\documentclass[%
 reprint,
superscriptaddress,
 amsmath,amssymb,
 aps,
 prl,
]{revtex4-2}

\usepackage{graphicx}
\usepackage{todonotes}
\usepackage{float}
\usepackage{mhchem}
\usepackage{dcolumn}
\usepackage{bm}
\usepackage{hyperref}
\hypersetup{
   breaklinks=true,   
   colorlinks=true,   
}

\begin{document}
\preprint{APS/123-QED}
\title{High-precision Penning trap mass measurements of neutron-rich chlorine isotopes at the $N = 28$ shell closure}%

\author{H.~Erington}%
\email{erington@frib.msu.edu}
\affiliation{Facility for Rare Isotope Beams, East Lansing, Michigan, 48824, USA}
\affiliation{Department of Physics and Astronomy, Michigan State University, East Lansing, Michigan 48824, USA}

\author{G.~Bollen}
\affiliation{Facility for Rare Isotope Beams, East Lansing, Michigan, 48824, USA}
\affiliation{Department of Physics and Astronomy, Michigan State University, East Lansing, Michigan 48824, USA}

\author{G.~Dykstra}
\affiliation{Department of Integrative Biology, Michigan State University, East Lansing, Michigan 48824, USA}

\author{A.~Hamaker}%
\affiliation{Facility for Rare Isotope Beams, East Lansing, Michigan, 48824, USA}
\affiliation{Department of Physics and Astronomy, Michigan State University, East Lansing, Michigan 48824, USA}

\author{C.~M.~Ireland}%
\affiliation{Facility for Rare Isotope Beams, East Lansing, Michigan, 48824, USA}
\affiliation{Department of Physics and Astronomy, Michigan State University, East Lansing, Michigan 48824, USA}

\author{C.~R.~Nicoloff}
\affiliation{Facility for Rare Isotope Beams, East Lansing, Michigan, 48824, USA}
\affiliation{Department of Physics and Astronomy, Michigan State University, East Lansing, Michigan 48824, USA}

\author{D.~Puentes}%
\affiliation{Facility for Rare Isotope Beams, East Lansing, Michigan, 48824, USA}
\affiliation{Department of Physics and Astronomy, Michigan State University, East Lansing, Michigan 48824, USA}

\author{R.~Ringle}
\affiliation{Facility for Rare Isotope Beams, East Lansing, Michigan, 48824, USA}
\affiliation{Department of Physics and Astronomy, Michigan State University, East Lansing, Michigan 48824, USA}

\author{S. Schwarz}
\affiliation{Facility for Rare Isotope Beams, East Lansing, Michigan, 48824, USA}

\author{C.~S.~Sumithrarachchi}
\affiliation{Facility for Rare Isotope Beams, East Lansing, Michigan, 48824, USA}

\author{A.~A.~Valverde}
\affiliation{Physics Division, Argonne National Laboratory, Lemont, Illinois 60439, USA}
\affiliation{Department of Physics and Astronomy, University of Manitoba, Winnipeg, Manitoba MB R3T 2N2, Canada}

\author{I.~T.~Yandow}%
\affiliation{Facility for Rare Isotope Beams, East Lansing, Michigan, 48824, USA}
\affiliation{Department of Physics and Astronomy, Michigan State University, East Lansing, Michigan 48824, USA}

\date{\today}

\begin{abstract}
\noindent Although it is known that the $N=28$ spherical shell closure erodes, the strength of the closure with decreasing proton number $Z<20$ is an open question in nuclear structure. In this region of interest, high-precision mass measurements of neutron-rich $^{43-45}$Cl isotopes were performed at the Low Energy Beam and Ion Trap (LEBIT) when coupled to the National Superconducting Cyclotron Lab. The resulting mass excesses (MEs) are ME($^{43}$Cl) = -24114.4(1.7) keV, ME($^{44}$Cl) = -20450.8(10.6) keV, and ME($^{45}$Cl) = -18240.1(3.7) keV, and improve the uncertainty of these masses by up to a factor of $\sim$40 compared to the previous values reported in the 2020 Atomic Mass Evaluation. Comparison to \textit{ab initio} calculations using the Valence-Space In-Medium Similarity Renormalization Group (VS-IMSRG) shows good agreement up to and including the closure.

\end{abstract}

\maketitle

\section{Introduction}

The nuclear shell model, with ``magic" numbers of increased stability, has been successful both theoretically and experimentally for decades. 
Originating from the introduction of the spin-orbit coupling term to the Woods-Saxon potential, magic numbers were explained by shell closures appearing at at 2, 8, 20, 28, 50, 82, and 126 across the nuclear chart \cite{SpinOrbit}. 
However, away from the valley of $\beta$-stability, we observe that magic numbers can shift, or their special properties fade \cite{MagicNumbers}. 
The experimental evidence for the erosion of the traditional magic numbers is key to our modern understanding of nuclear structure and the mechanisms that drive it \cite{SORLIN2008602}.

A particular region of interest is near the doubly magic nucleus $^{48}$Ca. One focus is quantifying the strength of the neutron shell closure at $N=28$ as the proton number, $Z$, decreases below 20, where there is the possibility for the erosion of the shell or the emergence of sub-shell closures. The magic number 28 is the first magic number produced from the spin-orbit interaction \cite{SpinOrbit}, allowing for insight into its evolution with neutron-proton asymmetry, as well as the contribution of other interaction terms. Furthermore, this region can be characterized in phenomenological shell-model calculations, offering quantifiable insight into the components of the nuclear interaction responsible for the structural evolution observed experimentally \cite{SMIRNOVA2010109}. Specifically, the recent valence-space formulation of the in-medium similarity renormalization group (VS-IMSRG) \textit{ab initio} calculations \cite{vsimsrg_recent} have shown agreement up to and at the drip line where experimental data are available. The VS-IMSRG and other models can be compared against new data in the $N=28$ region as it is obtained \cite{SpecComp1, SpecComp2, SpecComp3, SpecComp4, SpecComp5, Argon, Si_S_2025}.

From experimental studies of $E(2_1^+)$ values in the silicon chain ($Z=14$), it is known that the $N=28$ spherical shell gap is eroded around $^{42}$Si \cite{CollapseSi, DeformationSi, SPinSi, StructureSi}. There is also evidence of both shape and configuration coexistence in $^{44}$S \cite{GLASMACHER1997163, GammaNR, Shape_Sulphur, ProlateN28, Triple, NewCollectivitySulfur, IsomericCharacterSulfur}, which can indicate an ``island of inversion" \cite{Ni78SC}. 
Despite the data available, this region is not fully described \cite{EnergySpacingEvo, InverseKinematics, betaCl4647}.

To fully characterize the erosion in the $N=28$ region for $Z<20$, chains of intermediate proton number must be studied and compared to theory. The isotopic chains of interest are argon ($Z=18$), chlorine ($Z=17$), sulfur ($Z=16$), and phosphorus ($Z=15$). These lie between the doubly magic $^{48}$Ca and the spherical closure erosion occurring in $^{42}$Si. In addition to spectroscopic data, the strength of a closure is also characterized by the trends of nuclear binding energies along isotopic chains. Improving our knowledge of binding energies requires precise mass measurements that reduce the  uncertainty of present mass values and determine new masses far from $\beta$-stability.

Only two high-precision mass measurement results have been published in the $N=28$ region along the isotopic chains of interest \cite{Sulfur, Argon}. In 2009, the Low Energy Beam and Ion Trap (LEBIT) \cite{LEBIT}, when coupled to the National Superconducting Cyclotron Lab (NSCL), measured the masses of neutron-rich $^{40-44}$S isotopes \cite{Sulfur}. These measurements improved the characterization of the sulfur chain up to $N=28$ and are consistent with previous literature values; however, a large uncertainty remains for $ N=29$ and $30$. In 2020, high-precision mass measurements of $^{46-48}$Ar were performed with the ISOLTRAP mass spectrometer \cite{ISOLTRAP, MRTOF_ISOLTRAP} at ISOLDE/CERN \cite{Argon}. The results find a persistent but slightly reduced shell closure compared to the calcium chain; however, the binding energy strength must also be considered alongside spectroscopic data and theoretical calculations \cite{Nuance1, Nuance2}. Importantly, these argon measurements cover $N=28, 29, 30$, providing full characterization of the argon chain around $N=28$ for mass measurements. Other than these studies, only time-of-flight measurements have reached beyond $N=28$ for all isotopic chains of interest, which have experimental uncertainties greater than 100 keV \cite{SC_Review2000, JURADO200743, NSCL_TOFmeas} that cannot reveal the closure's strength. For this reason, precision mass measurements in this region of the nuclear chart are highly desirable to fully characterize the strength of the shell closure.

This work reports the first high-precision mass measurements of $^{43-45}$Cl using the Time-of-Flight Ion Cyclotron Resonance (TOF-ICR) technique \cite{ToF1, ToF2, ToF3} at LEBIT. These measurements provide a characterization of the mass surface up to $N=28$, and the values are compared to previous measurements. Additionally, the new binding-energy trends are compared to the recent VS-IMSRG calculations to test agreement at $N=28$.

\section{Experiment}

At the NSCL’s Coupled Cyclotron Facility, a $^{48}$Ca primary beam was accelerated to 140 MeV/u and impinged on a $^9$Be target of thickness 846 mg/cm$^2$. The produced fragments were sent through the A1900 fragment separator \cite{A1900} where the $^{43-45}$Cl isotopes were identified at the focal plane via the energy loss and time-of-flight ($\Delta E$ vs TOF) PID method and selected. After the separation, the beam proceeded via a momentum compression beamline to the gas stopping area. Before entering a gas cell, the beam was slowed with aluminum degraders and dispersion matched with an aluminum wedge. The effective thicknesses of the rotatable degraders were adjusted for each individual chlorine isotope to optimize stopping efficiency, allowing the beam to enter the gas cell at less than 1 MeV. The gas stopping area utilizes two gas cells, the Room Temperature Gas Cell (RTGC) \cite{SUMITHRARACHCHI2020305} and the Advanced Cryogenic Gas Cell (ACGS) \cite{LUND2020378}. The gas cell used for each chlorine isotope was chosen based on the impurities and efficiency upon extraction. 

In the RTGC or ACGS, the highly charged ions undergo collisions with the helium gas and the chlorine isotopes recombined down to a charge state of +1. In the RTGC the ions are transported by a combination of radiofrequency (RF) and direct current (DC) fields and gas flow. In the ACGS, the ions are transported via ion-carpet surfing \cite{IonSurfing} using an RF electric wave and a DC push field. In both gas cells, the ions are then extracted into an RF quadrupole (RFQ) ion guide. The gas stopping and LEBIT facilities are raised to 30 kV while the transport beam line between them is at ground potential. This configuration allows the ions to be accelerated to 30 keV during transport and then slowed before entering LEBIT. During transport, the ions are sent through a dipole magnet with a resolving power of approximately 1500, which selected all ion species based on the mass-to-charge ratio $A$/Q for the corresponding chlorine isotope, ensuring that bare $^{43}$Cl$^+$, $^{44}$Cl$^+$, and $^{45}$Cl$^+$ was delivered to LEBIT.

\begin{figure}[t]
    \centering
    \includegraphics[width=\linewidth]{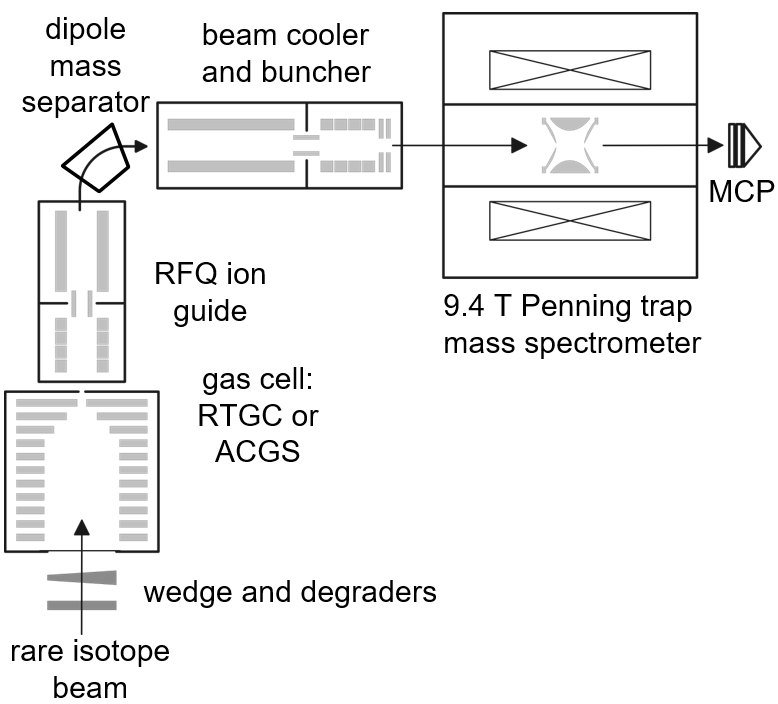}
    \caption{Schematic of the NSCL gas cell and LEBIT facility.}
    \label{fig:NSCL_schematic}
\end{figure}

After entering LEBIT, the selected continuous beam was injected into a linear buffer-gas-filled Paul trap cooler and buncher \cite{CoolerBuncher}. Here, the beam is accumulated for a set duration, cooled with a He buffer gas, and then ejected in distinct ion bunches. Following extraction, the ion bunches are guided and injected into the 9.4 T Penning trap mass spectrometer \cite{PenningTrap}. Figure \ref{fig:NSCL_schematic} shows a schematic of the gas cell and LEBIT facility.

Ions in the Penning trap are confined radially via a homogeneous magnetic field $B$ and axially via a quadrupolar electrostatic field. While confined, the ions undergo three independent eigenmodes of motion resulting from the superposition of the two fields. Two of these occur in the radial plane of motion, $\nu_+$ (reduced cyclotron frequency) and $\nu_-$ (magnetron frequency), and one in the axial plane, $\nu_z$ (axial frequency). Under typical operating conditions where the magnetic field is much stronger than the electrostatic field, $\nu_- \ll \nu_z < \nu_+$, and in an ideal trap, the approximation of the cyclotron frequency, $\nu_c$, can be written as 

\begin{equation}
    \nu_c = \nu_- + \nu_+ = \frac{qB}{2\pi m}
\end{equation}

\noindent where $q$ is the charge of the ion, $B$ is the strength of the magnetic field, and $m$ is the mass of the ion. The error resulting from this approximation is negligible compared to the statistical errors \cite{SidebandMassSpecApprox} obtained in these measurements.

\begin{figure}[t]
    \centering
    \includegraphics[width=\linewidth]{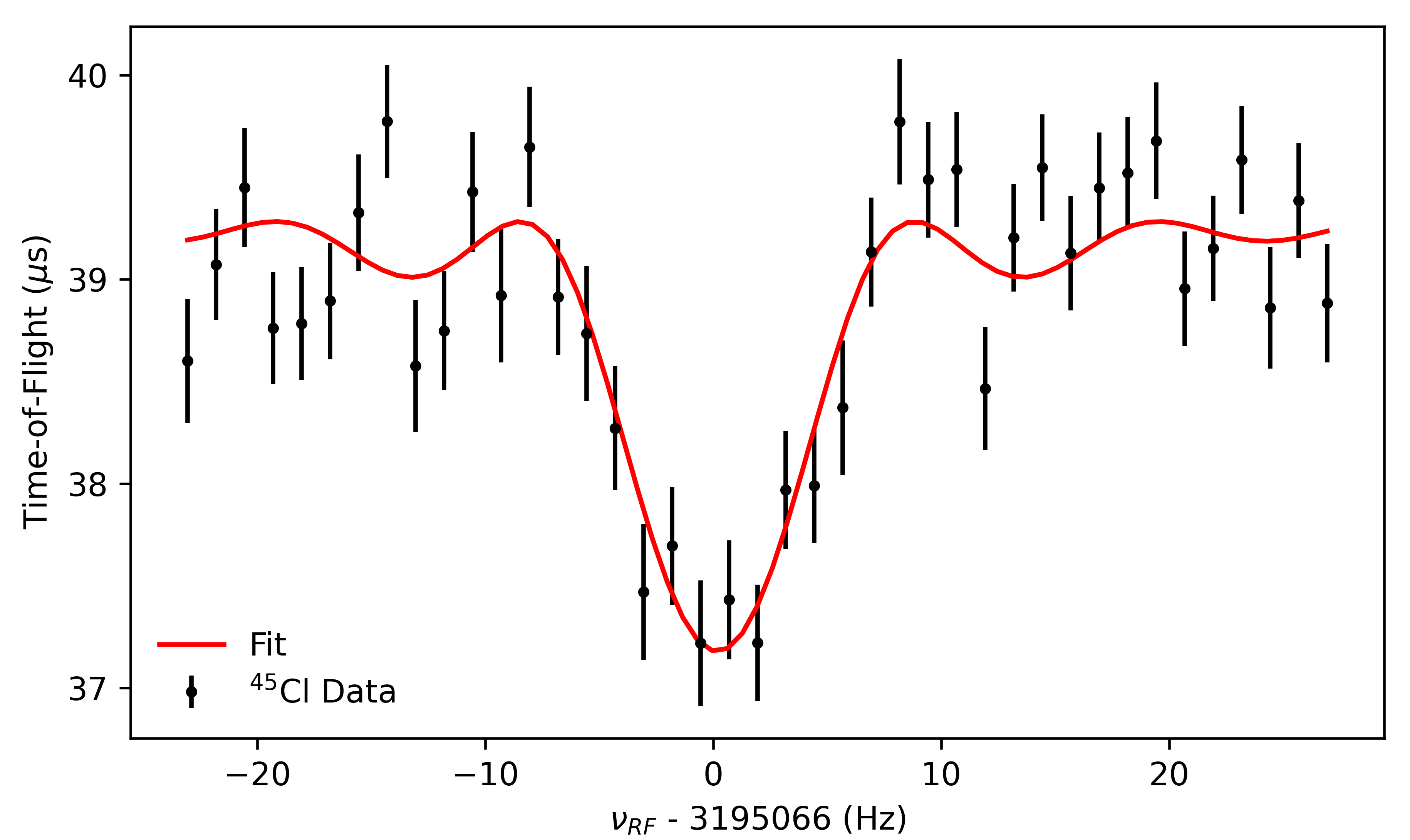}
    \caption{A summed TOF spectrum for all of the $^{45}$Cl measurements taken using the LEBIT 9.4 T Penning trap mass spectrometer. The spectrum is formed by 3789 ions. The width of the central dip corresponds to the inverse of the excitation time of $t_{RF} =$ 100 ms. A $\chi^{2}$ minimization fit to the analytical curve described in \cite{ToF3}, depicted in red, was used to determine the frequency $\nu_{RF} = \nu_{c}$ occurring at the minimum time-of-flight.}
    \label{fig:TOFcurve}
\end{figure}

The masses of $^{43-45}$Cl were determined by the Time-of-Flight Ion Cyclotron Resonance (TOF-ICR) technique. Ions extracted from the cooler-buncher were steered off-axis with Lorentz steerers \cite{LorentzSteerer}, inducing initial magnetron motion upon capture in the trap. An azimuthal quadrupolar RF field pulse $\nu_{RF}$ close to the expected cyclotron frequency is applied to the ions for a chosen excitation time, $t_{RF}$. In the case where $\nu_{RF} = \nu_c$, the slow magnetron motion is fully converted into the fast reduced cyclotron motion, increasing the radial kinetic energy of the ions. Upon ejection from the trap, ions move through the magnetic field gradient before striking the Multi-Channel Plate (MCP) detector. Ions with larger radial kinetic energy experience a greater force when moving through the magnetic field gradient due to their interaction with the magnetic moment, causing them to arrive at the MCP detector faster and, therefore, producing a shorter time of flight (TOF). Repeating the cycle of trapping, excitation, and measurement at different RF frequencies results in a resonance curve with a minimum at $\nu_{RF} = \nu_c$. An example of this resonance curve with the summed $^{45}$Cl data is shown in Fig.~\ref{fig:TOFcurve}.

A reference ion is needed to obtain the mass of the ion of interest. This reference enables the mass of the ion of interest to be determined relative to the reference mass via the cyclotron frequency ratio $R$,

\begin{equation}
    R = \frac{\nu_c}{\nu_{c, ref}} = \frac{q \cdot m_{ref}}{q_{ref} \cdot m}
\end{equation}

\noindent where $q$ and $q_{\mathrm{ref}}$ are the charges and $m$ and $m_{\mathrm{ref}}$ are the masses of the ion of interest and the reference ion, respectively.

For the case when both the reference ion and the ion of interest are singly charged, the final reported atomic mass $M$ is found using 

\begin{equation} \label{eq: 3}
    M = ([M_{ref} - m_e]/\bar{R}) + m_e
\end{equation}

\noindent where $\bar{R}$ is the weighted average of all cyclotron frequency ratios, $M_{ref}$ is the atomic mass of the neutral reference atom/molecule and $m_e$ is the electron mass. Electron binding energy is neglected as the statistical uncertainty of the measurement dominates. All measurements reported were performed with singly charged ions.

For all measurements, reference ion measurements are interleaved with those of the ion of interest to obtain a linearly interpolated value of
$B$ at the time the ion of interest is measured. Seven independent measurements with $t_{RF}=100$ ms were made of $^{43}$Cl$^+$ using $^{39}$K$^+$ as the reference, of which the final three utilized the pulsed Ramsey resonance technique \cite{RamseyTechnique}. Seven independent measurements with $t_{RF}=50$ ms were made of $^{44}$Cl$^+$, using [$^{12}$C$^{14}$N$^{1}$H$_2$$^{16}$O]$^+$ ($A=44$) and $^{39}$K$^+$ as references. Four independent measurements with $t_{RF}=100$ ms of $^{45}$Cl$^+$ were made using [$^{28}$Si$^{16}$O$^{1}$H]$^+$ ($A=45$) as the reference ion. The molecules used as references were produced in the gas cell due to ionization of impurities in helium gas and were co-delivered with the ion of interest. The constituent masses of such molecules are known with high precision, resulting in a well-defined cyclotron frequency that enabled their identification and subsequent use as reference species. Each measurement required between 20 min and 2 hr to produce a resonance curve, depending on the isotope.

\section{Results}

\begin{figure}[H]
    \centering
    \includegraphics[width=\linewidth]{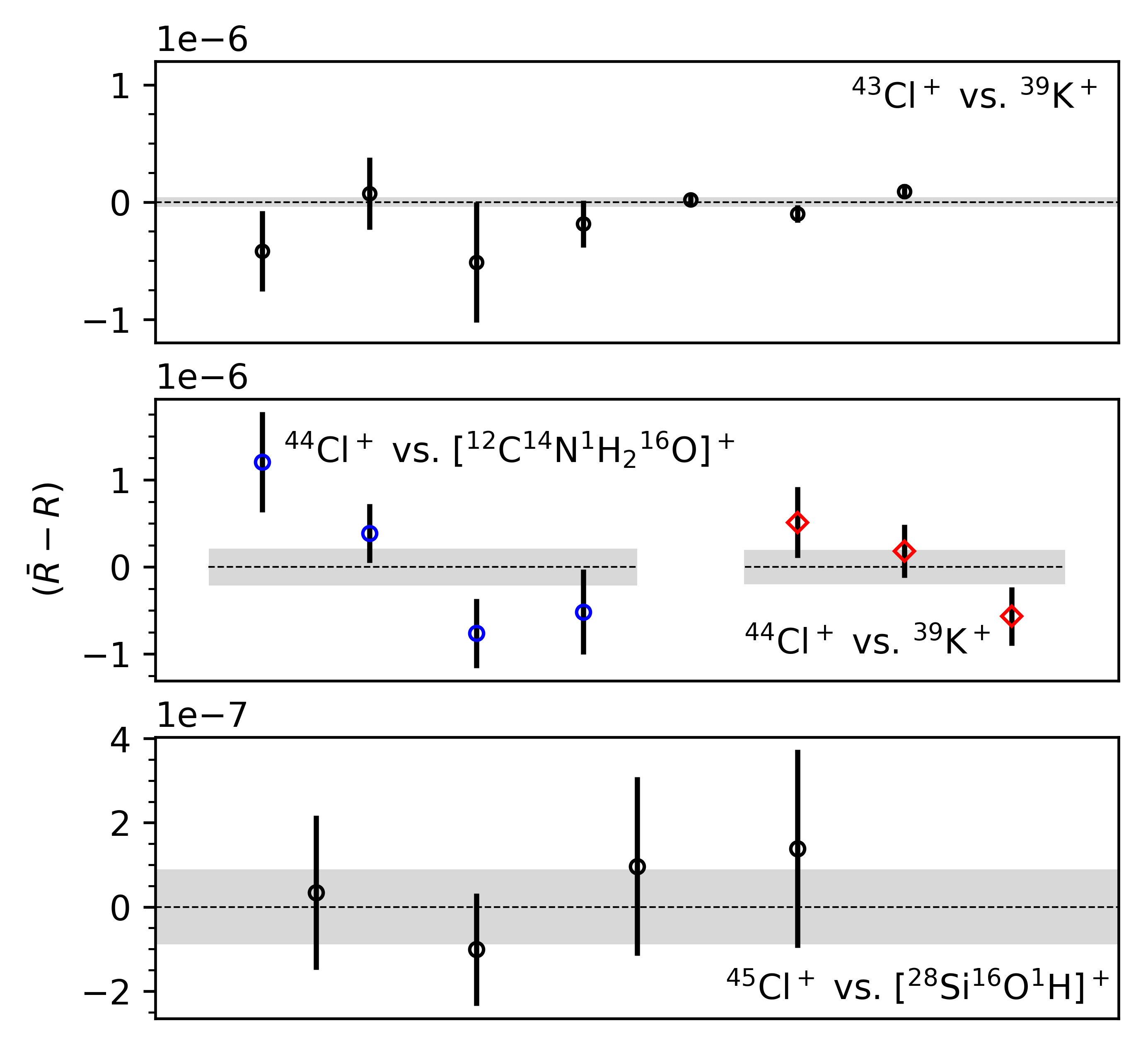}
    \caption{Difference of the measured frequency ratio $R$ from the mean ratio $\bar{R}$ for each independent series of measurements. The data points are grouped by the reference ion, as indicated. Gray band shows the $\pm 1 \sigma$ uncertainty in $\bar{R}$.}
    \label{fig:all-Cl-measurements}
\end{figure}

The resulting difference of the frequency ratio for each individual measurement, $R$, from the mean frequency ratio, $\bar{R}$, for the series of measurements is shown in Fig.~\ref{fig:all-Cl-measurements}.

{\renewcommand{\arraystretch}{1.2}
\begin{table*}[t]
\caption{\label{tab:table1} Mean ratios $\bar{R}$ and the corresponding mass excesses calculated from the data shown in Fig.~\ref{fig:all-Cl-measurements}. Values from the AME2020 \cite{AME2020} are shown for comparison. Half-lives are taken from the NUBASE2020 evaluation \cite{Nubase2020}.}
\begin{ruledtabular}
\begin{tabular}{lccccc}
 \multicolumn{2}{c}{} & \multicolumn{2}{c}{} & \multicolumn{2}{c}{Mass Excess (keV)}\\ \cline{5-6}
 
 Ion of Interest & Half-life & Reference & $\bar{R} = \nu_c/\nu_{c, ref}$ & This work & AME2020 \\ \hline

 & & & & & \\
 
 $^{43}$Cl$^{+}$ & 3.13(9) s & $^{39}$K$^{+}$ & 0.906677379(40) & -24114.4(1.7) & -24160(60) \\
 
 & & & & & \\
 
 $^{44}$Cl$^{+}$ & 562(106) ms
 & [$^{12}$C$^{14}$N$^{1}$H$_2 ^{16}$O]$^+$ & 1.00080941(39) & -20453.0(16.1) &  \\
 
 &  & $^{39}$K$^{+}$ & 0.88597939(30) & -20448.6(14.0) &  \\
&  &  & Average & -20450.8(10.6) & -20480(90) \\
 & & & & & \\
 
 $^{45}$Cl$^{+}$ & 413(25) ms & [$^{28}$Si$^{16}$O$^{1}$H]$^+$ & 0.999983276(89) & -18240.1(3.7) & -18260(140) \\
\end{tabular}
\end{ruledtabular}
\end{table*}}

Several sources of systematic effects contribute to the uncertainty $\delta_R$ in $\bar{R}$. When $^{39}$K$^+$ is used as the reference ion, there is a mass difference of a few mass units between it and the ion of interest. It has been found that systematic shifts in $\bar{R}$ (Eq.~\ref{eq: 3}) scale linearly with this mass difference, which is over an order of magnitude smaller than the statistical uncertainty. There are also mass-dependent shifts related to inhomogeneity in the magnetic field and trap imperfections, which have been studied at LEBIT in detail and are known to add $\delta_R \approx 2 \times 10^{-10}/u$ \cite{SysShift_BfieldandTrap}, as well as non-linear temporal shifts which contribute $\delta_R < 10^{-9}$ per hour \cite{SysShift_NLtemporal}. To mitigate these, regular reference measurements were conducted. Isobaric contamination was mitigated using a dipolar RF excitation applied to the central ring electrode near their respective reduced cyclotron frequencies \cite{DipoleCleaning}. Finally, events with greater than six detected ions were discarded to avoid shifts from Coulomb interactions in the trap for all measurements.

\begin{figure}[h]
    \centering
    \includegraphics[width=\linewidth]{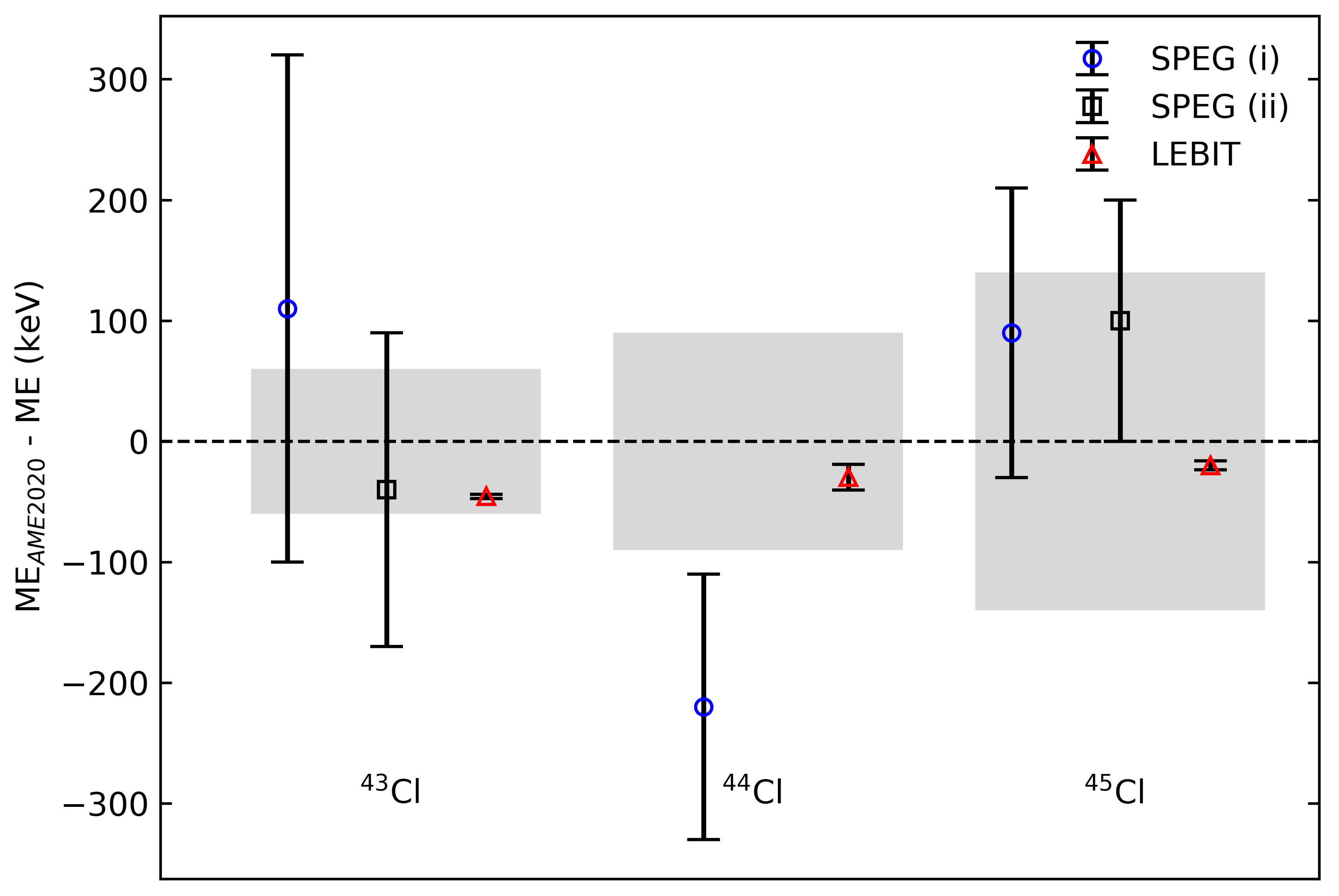}
    \caption{Comparison between the values found for the $^{43-45}$Cl mass excesses obtained in this work (red triangle) and previous works subtracting the experimental mass excess from the AME2020 mass excess. The gray band represents the AME2020 one standard deviation.}
    \label{fig:ME_comparison}
\end{figure}

The mean frequency ratios $\bar{R}$ for each independent series of measurements are listed in Table \ref{tab:table1}, along with their statistical uncertainties. Additionally, Fig.~\ref{fig:ME_comparison} shows the results compared to both the Atomic Mass Evaluation 2020 (AME2020) \cite{AME2020} and the previous measurements of these isotopes \cite{SC_Review2000, JURADO200743}. Each measured isotope is less bound by 45.5(60.0) keV, 29.5(90.6) keV, and 19.8(140.0) keV than the AME2020 value, with an increased precision by a factor of 34, 8, and 37 for $^{43}$Cl, $^{44}$Cl, and $^{45}$Cl respectively.

\section{Discussion}

The mass values obtained in this work agree well with the AME2020 and are more precise, see Table~\ref{tab:table1}. Using these newly obtained masses, an evaluation of the neutron shell gap up to $N=28$ can be performed. In this evaluation, quantities known as mass filters are of interest, determined via mass differences. One such quantity is the three-point estimator of the pairing gap, defined as 

\begin{equation}
    \begin{split}
        \Delta_{3n}(N,Z) & = \frac{(-1)^N}{2} [\text{ME}(N+1, Z) \\
        & \qquad - 2\text{ME}(N, Z) + \text{ME}(N-1, Z)]
    \end{split}
\end{equation}

\noindent where ME denotes the mass excess of the respective isotope. This quantity is most commonly used to show the odd-even staggering effect, which is produced by the energy discrepancy between even and odd numbered nuclei due to the presence or absence of paired nuclei. At the crossing of a shell closure, this staggering is enhanced, as the $\Delta_{3n}$ is also related to the single-particle level spacing \cite{OES_D3n}. It is related to the one-neutron shell gap via 

\begin{equation} \label{eq: 5}
    \Delta_{1n}(N_{\text{closure}}, Z) = 2 \times \Delta_{3n}(N_{\text{closure}}, Z).
\end{equation}

\begin{figure}[h]
    \centering
    \includegraphics[width=\linewidth]{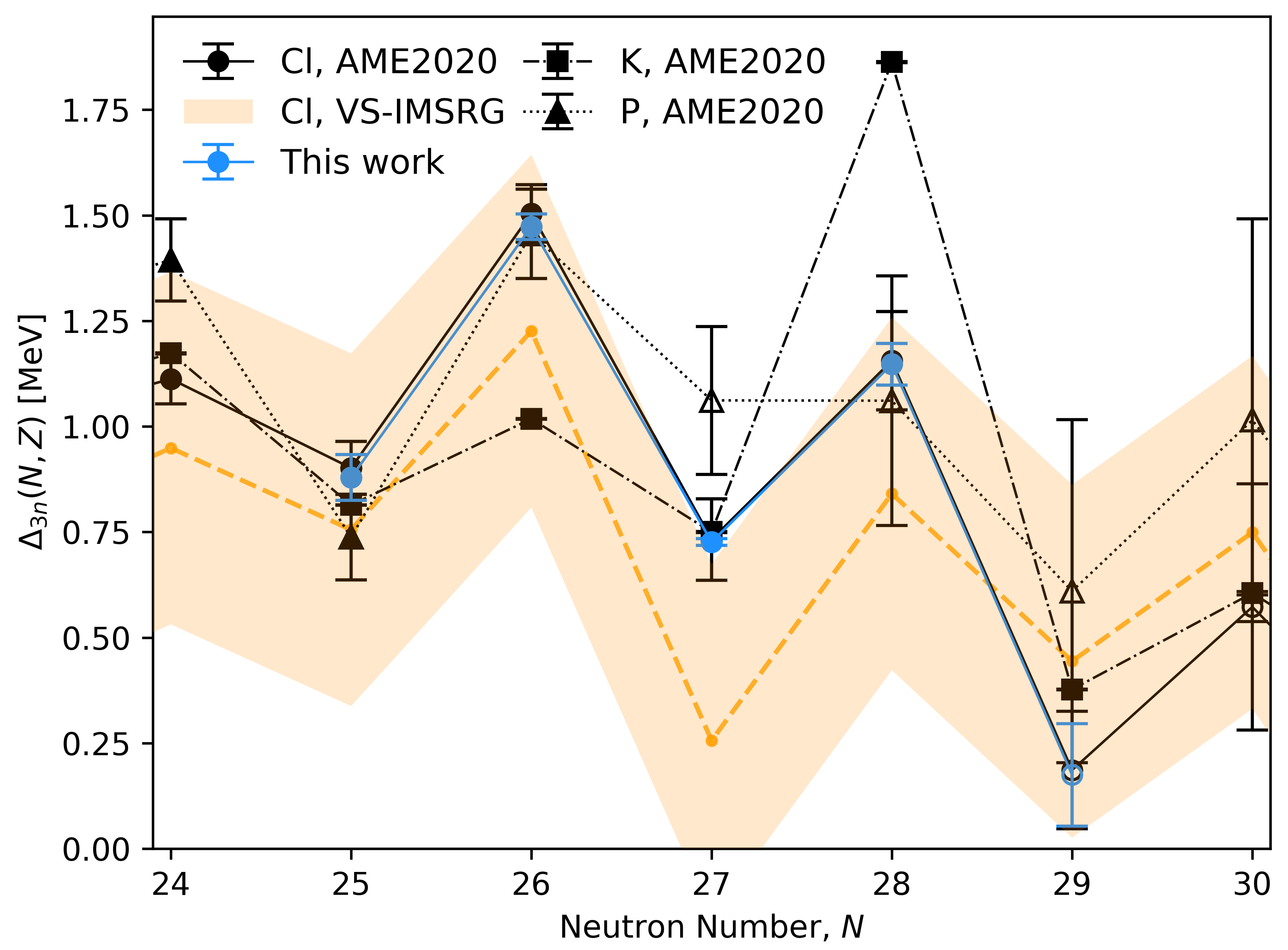}
    \caption{The three-point estimator of the pairing gap ranging from $N=24$ to $N=30$. Potassium ($Z=19$) and phosphorus ($Z=15$) values are extracted from the AME2020 and are represented by squares and triangles, respectively. Black circles represent chlorine ($Z=17$) values from the AME2020, and the calculated values impacted by this work are shown as blue circles. All calculated values that include AME2020 extrapolations are indicated by open shapes. VS-IMSRG values and uncertainties are indicated by orange dots and the shaded orange region.}
    \label{fig:d3n}
\end{figure}

From this relation, one would expect a significant enhancement at $N=28$. Figure \ref{fig:d3n} shows the $\Delta_{3n}$ for the potassium, chlorine, and phosphorus chains. The masses measured in this work contribute to the calculated values at $N=25$ through $N=29$ for the chlorine chain. The plot shows a significant decrease in the one-neutron shell gap for chlorine ($\Delta_{1n}(28,17) = 2.294 \pm 0.098$ MeV) compared to potassium ($\Delta_{1n}(28,19) = 3.726 \pm 0.002$ MeV), while maintaining a slightly greater one-neutron shell gap compared to phosphorus ($\Delta_{1n}(28,15) = 2.122 \pm 0.591$ MeV). This is expected as the proton number decreases, however, it should be noted that the chlorine value at $N=28$ still requires the use of the AME2020 value for $^{46}$Cl. Due to the large uncertainty remaining for both chlorine and phosphorus, the difference in their $\Delta_{1n}$ values may be more or less pronounced.

Additionally, comparing to theoretical models utilizing \textit{ab initio} methods and interactions developed in chiral effective field theory \cite{heiko-review} in this region is critical considering that binding energy observables validate where these models are successful and where additional insight can be gained. Using the VS-IMSRG \cite{vsimsrg1, vsimsrg2, vsimsrg3, HERGERT2016165}, agreement between theory and experiment was found for the argon chain in the $N=28$ region \cite{Argon, HERGERT2016165}. 

From updated VS-IMSRG calculations that use the 1.8/2.0(EM) NN+3N Hamiltonian and include a correction for systematic error through Bayesian analysis \cite{vsimsrg_recent}, the $S_{1n}$ values and their errors were used to calculate the theoretical $\Delta_{3n}$ values for the chlorine chain based on Eq.~\ref{eq: 5}, shown as the shaded region in Fig.~\ref{fig:d3n}. The errors shown in this shaded region are propagated from the 68\% uncertainty band of the posterior distribution for the $S_{1n}$ values outlined in \cite{vsimsrg_recent}. The plot shows that the trend in the calculation from the experimental data is consistent with the VS-IMSRG calculations, albeit with some offset within error, until the point at $N=29$, where the error bar remains large due to the dominant contribution of the mass uncertainty for $^{46}$Cl and $^{47}$Cl. If confirmed, this discrepancy in the trend could provide additional insight into pairing effects and the mean field contribution \cite{JURADO200743, OES_D3n}. As such, precise mass measurements beyond $N=28$ would be required to better characterize the shape of the three-point estimator and the strength of the shell closure for the chlorine chain.

\section{Conclusions}

Penning trap mass measurements of $^{43-45}$Cl were performed at LEBIT, improving the uncertainty of these isotopes by up to a factor of $\sim$40. The results are in agreement with previous values from the AME2020 and enable more precise determinations of nuclear structure effects using mass filters. Theoretical calculations and measurement calculations of the three-point estimator agree reasonably well. However, the large uncertainty from the less precise and unmeasured masses from AME2020 does not yet allow  the $N=28$ shell closure strength to be determined with desirable accuracy. High-precision mass measurements of $^{46,47}$Cl, as well as measurements in the S and P isotope chains beyond $N=28$, would bring clarity to this question. As of 2022, LEBIT is now coupled to the Facility for Rare Isotope Beams (FRIB), which can provide the necessary rates for high-precision measurements beyond $N=28$ and down the isotopic chains of interest, making them within reach in the near future.

\section{Acknowledgments}

This material is based upon work supported by the U.S. Department of Energy, Office of Science, Office of Nuclear Physics, and used resources of the Facility for Rare Isotope Beams (FRIB) Operations, which is a DOE Office of Science User Facility under Award Number DE-SC0023633. This work was conducted with the support of Michigan State University and the U.S. National Science Foundation under contracts nos. PHY-1565546 and PHY-2111185, the DOE, Office of Nuclear Physics under contract no. DE-AC02-06CH11357, DE-AC02-05CH11231, and DE-SC0022538. A.A.V acknowledges support from the DOE, Office of Nuclear Physics under contract no. DE-AC02-06CH11357 and  NSERC (Canada), Application No. SAPPJ-2018-00028. C.M.I. acknowledges support from the ASET Traineeship under the DOE award no. DE-SC0018362.

\bibliography{refs}

\end{document}